\documentclass[prd,superscriptaddress,showpacs,showkeys,notitlepage,floatfix]{revtex4-1}

\usepackage[utf8]{inputenc}
\usepackage{amsmath,amsfonts,amsthm,amssymb}

\usepackage{graphicx}

\usepackage[colorlinks=true,
            linkcolor=blue,
            urlcolor=blue,
            citecolor=blue]{hyperref}

\begin{document}
\title{$D=11$ cosmologies with teleparallel structure}

\author{Christian G. B\"{o}hmer}
\email{c.boehmer@ucl.ac.uk} 
\affiliation{Department of Mathematics,
University College London,
Gower Street,
London, WC1E 6BT,
United Kingdom}

\author{Franco Fiorini}
\email{francof@cab.cnea.gov.ar.} 
\affiliation{Departamento de Ingeniería en Telecomunicaciones and Instituto Balseiro, Centro Atómico Bariloche (CONICET), 
Av. Ezequiel Bustillo 9500, 
CP8400, S. C. de Bariloche, 
Río Negro, 
Argentina.}

\author{P. A. Gonz\'{a}lez}
\email{pablo.gonzalez@udp.cl} 
\affiliation{Facultad de Ingenier\'{\i}a y Ciencias, 
Universidad Diego Portales, 
Avenida Ej\'{e}rcito Libertador 441, 
Casilla 298-V, Santiago, 
Chile.}

\author{Yerko V\'{a}squez}
\email{yvasquez@userena.cl}
\affiliation{Departamento de F\'isica y Astronom\'ia, 
Facultad de Ciencias, Universidad de La Serena,
Avenida Cisternas 1200, La Serena, 
Chile.}

\date{\today }

\begin{abstract}

The presence of additional compact dimensions in cosmological models is studied in the context of modified teleparallel theories of gravity. We focus the analysis on eleven dimensional spacetimes, where the seven dimensional extra dimensions are compactified. In particular, and due to the importance that global vector fields play within the conceptual body of teleparallel modified gravity models, we consider the additional dimensions to be products of parallelizable spheres. The global vector fields characterizing the different topologies are obtained, as well as the equations of motion associated to them. Using global dynamical system techniques, we discuss some physical consequences arising because of the existence of the extra dimensions. In particular, the possibility of having an early inflationary epoch driven by the presence of extra dimensions without other matter sources is discussed. 

\end{abstract}
\maketitle

\section{Introduction}
\label{sec:intro}

Despite being somewhat counterintuitive, the possible existence of extra spatial dimensions has a distinguished history in theoretical physics and can be considered a well established idea by now. Since their introduction in the early 1920s \cite{Kaluza:1921tu} as a tentative approach of unifying electrodynamics and gravitation under a common geometrical context, there has been a considerable and growing interest in physical models involving more dimensions than the three spatial dimensions and time which seem to govern our daily experience. These interests rapidly went far beyond the unifying purposes present in the original models, for it was demonstrated that the inclusion of extra dimensions could solve several long standing problems in theoretical physics. For instance, it was argued that by extending the number of dimensions, two of the most important \emph{hierarchy} problems could find an elegant resolution; the Higgs mass hierarchy problem \cite{Arkani,Randall} and the problem of the cosmological constant \cite{Dvali}. Another area in which the existence of extra dimensions seems to play an important role is quantum gravity. Even though no general consensus exist towards the formulation of a consistent quantum theory of the gravitational field, it is well know that one of the candidates, string theory (M-theory), can be consistently constructed in spaces with extra six (seven) spatial dimensions \cite{Green,Witten}.

At a purely gravitational level, the study of theories including (or formulated on) manifolds with additional spatial dimensions has been worked since  Lovelock's expansion was discovered \cite{Lovelock} (see also \cite{Maestro}). It emphasized the fact that General Relativity (GR) does not seem to be the most natural theory of gravity when the number $D$ of spacetime dimensions is bigger than four. According to the original philosophy surrounding GR, if we remain within the metric description of gravity, Lovelock's Lagrangian is the only one assuring second order field equations which are automatically conserved. If $D>4$ the expansion necessarily contains higher order terms in the curvature; for instance, in five and six spacetime dimensions, the Lagrangian density is not just the Hilbert-Einstein term, but it contains a specific quadratic combination of curvature terms given by the Gauss-Bonnet term. This quadratic piece is `harmless' in $D=4$ in the sense that it is a topological invariant, the Euler density, which does not contribute to the field equations. However, when $D>4$ the Gauss-Bonnet term not only becomes dynamical, it also arises as the curvature correction to the Einstein-Hilbert Lagrangian coming from supersymmetric string theory \cite{Deser}. This seems to indicate that the study of classical gravitation in extra dimensions is well motivated from a theoretical point of view. 

Research of the previous decades has contrived a number of modifications and extensions of Einstein's original theory, and many are being studied extensively in a variety of different contexts. Among the many theories developed, the so called $f(T)$-gravity or modified teleparallel gravity has attracted much attention in recent years. Originally \cite{Noso}, models of this type were proposed as a high energy modification of GR in relation to the existence of strong curvature singularities in cosmological models. It became clear soon after that $f(T)$ theories exhibit interesting late time cosmological implications as well, as witnessed in the study of missing matter problems, and concerning the current acceleration stage of the universe without introducing any exotic matter content \cite{Bengochea:2008gz}--\cite{f(T)cosmo0p}. Recent developments on $f(T)$-gravity concerning cosmological implications can be found in \cite{f(T)cosmo1}--\cite{f(T)cosmo5}, for instance.

The subject of extra dimensions in $f(T)$ gravity was scarcely presented in the literature \cite{Capozziello:2012zj}--\cite{C-Q4}, see also Refs.~ \cite{Gonzalez:2015sha,Gonzalez:2019tky}. The reason for this relative absence of contributions in the area is clearly understood when the structure of the $f(T)$ field equations is considered more carefully. Unlike many of the modified gravity theories in vogue, $f(T)$ gravity is formulated in Weitzenb\"{o}ck spacetime which is characterized by a set of 1-forms $E^{a}(x)=E^{a}_{\mu}dx^{\mu}$ producing torsion instead of curvature; the vielbein field $E^{a}(x)$, which encodes the dynamics of the gravitational field, determine the structure of the spacetime by means of a parallelization process. This means, among other things, that the field equations are not locally Lorentz invariant, or at least they are not in the usual sense \cite{Krssak:2018ywd}, an issue that is not fully settled yet. Of course, $f(T)$ modified gravity includes GR as a limit when $f(T)=T$, in which case one speaks of the teleparallel equivalent of GR. Therefore the Lorentz covariance is fully restored in those regimes or scales where the gravitational field is correctly described by GR. However, near spacetime singularities, for example, the structure of the fields $E^{a}(x)$ is fixed only up to a certain subgroup of the Lorentz group which is characteristic of the spacetime under consideration. This poses an additional technical complication at the time of solving the field equations. The field equations are second order partial differential equations which determine the full components of the vielbein $E^{a}(x)$, and not only those associated to the metric tensor $g_{\mu\nu}=E^{a}_{\mu}E^{b}_{\nu}\eta_{ab}$.  

In \cite{Fiorini:2013hva} we started the program of characterizing multidimensional cosmological models with $f(T)$ structure by assuming that the compact extra dimensions consisted of topological products of spheres. We did so by considering cosmological manifolds of the form $\mathcal{M}=\mathrm{FRLW}_{4} \times \mathcal{M}_{in}$, where $\mathrm{FLRW}_{4}$ represents the four dimensional, spatially flat Friedmann-Lema\^\i tre-Robertson-Walker spaces which are consistent with our present day understanding of the large scale structure of the universe. In \cite{Fiorini:2013hva} we dealt with cases up to $D=7$ where the spatial extra dimensions $\mathcal{M}_{in}$ were products of spheres up to dimension three. 

In this paper, in turn, we extend the analysis by focusing on $D=11$, and considering $\mathcal{M}_{in}$ as a seven dimensional manifold constructed out of products of parallelizable spheres. This restriction of considering only the parallelizable spheres simplifies the subsequent analysis as there are only three such spheres $S^{1}$, $S^{3}$ and $S^{7}$. On one hand, the number $n_{s}$ of different products of arbitrary spheres grows rapidly as the number $N$ of extra dimensions increases; we actually have $n_{s}=P(N)$, the partition of $N$. In the general case, this makes the problem hard to deal with for $D=11$, where one would have $n_{s}=P(7)=15$ distinct cases to deal with. On the other hand, in view of the close relationship between the 1-form fields $E^{a}(x)$ and the parallelizations underlying a given manifold, the structure of $E^{a}(x)$ turns out to be easier to deal with when $\mathcal{M}_{in}$ is itself a product of parallelizable submanifolds; the $E^{a}(x)$ simply inherit the product structure of $\mathcal{M}_{in}$. 

The paper is organized as follows: In section \ref{sec:fundamental} we present a concise account on $f(T)$ gravity. Section \ref{sec:spherical} is devoted to the structure of $E^{a}(x)$ for the four different cases under consideration arising by imposing the parallelizability condition on every member of the product in $\mathcal{M}_{in}$. Albeit technical and cumbersome, this section is crucial for finding the proper set of field equations in every case, which can be seen as the main contribution of this work. Some consequences of the field equations are obtained in Section \ref{dynamic}, where a global dynamical system analysis is performed on one of the cases exposed. More specifically Section \ref{sec:spherical} discusses the case $\mathcal{M}_{in}=S^7$. Finally, we discuss our results in Section \ref{sec:conclution}. Various Appendices are required to present some of the lengthy equations and certain technical details which are not essential in the main body. Appendix \ref{remnantgroup} briefly discusses the remnant symmetries, which are related to the different parallelizations admitted by a given manifold. Appendix \ref{anotherexample} contains a further example in which $\mathcal{M}_{in}$ is not constructed as a product of parallelizable spheres. 

Throughout the paper, we will adopt the signature $(+,-,-,-,\ldots,-)$, Latin indices $a:(0),(1),\ldots$ in $E^{a}_{\mu}(x)$ refer to tangent-space objects while Greek $\mu:0,1,\ldots$ denote spacetime indices. Dual vector basis $e_{a}=e_{a}^{\mu}\partial_{\mu}$ are defined according to $E^{a}_{\mu}e_{a}^{\nu}=\delta_{\mu}^{\nu}$ and $E^{a}_{\mu}e_{b}^{\mu}=\delta_{b}^{a}$.

\section{Brief notes on modified teleparallel gravity}
\label{sec:fundamental}

The extended gravitational schemes with absolute parallelism, often referred to as $f(T)$ theories, take as a starting point the Teleparallel Equivalent of General
Relativity (TEGR), see for instance \cite{Maluf:2013gaa,TEGRbook}. We will summarize here the basic elements needed to present the key ideas required for the formulation of our work. For a thorough introduction to $f(T)$ gravity as well as its mathematical foundations the reader is referred to \cite{Nos3,RyM}, for instance.

The spirit underlying TEGR find its motivation in the equivalence between the Riemann and Weitzenb\"{o}ck
formulations of GR, which can be summarized in the equation
\begin{equation}
    T = -R + 2\; e^{-1}\;\partial _{\nu }(e\,T_{\sigma }{}^{\sigma \nu}\,) \,,
    \label{divergence}
\end{equation}
where $e=\det(e_{a}^{\mu})=\sqrt{\det(g_{\mu\nu})}$. On the left-hand side of Eq. (\ref{divergence}) we have the so-called Weitzenb\"{o}ck invariant
\begin{equation}
    T = S^{\rho }{}_{\mu \nu }\ T_{\rho }{}^{\mu \nu } \,,
    \label{Weitinvar}
\end{equation}
where $T_{\rho }{}^{\mu \nu}$ are the local (spacetime) components of the torsion two-form
$T^{a}=de^{a} = \frac{1}{2} T^a{}_{\mu\nu} dx^\mu \wedge dx^\nu$ coming from the Weitzenb\"{o}ck connection $\Gamma^\lambda_{\nu%
\mu}=\,e_{a}^{\lambda }\,\partial _{\nu }e_{\mu }^{a}$, and $S_{\rho \lambda \mu}$ is defined according to
\begin{equation}
    S^{\rho }{}_{\mu \nu} = \frac{1}{4}\,(T^{\rho }{}_{\mu \nu }-T_{\mu \nu }{}^{\rho }+
    T_{\nu \mu }{}^{\rho})+\frac{1}{2}\ \delta _{\mu }^{\rho }\
    T_{\sigma \nu }{}^{\sigma }-\frac{1}{2}\ \delta _{\nu }^{\rho }\
    T_{\sigma \mu }{}^{\sigma }.
    \label{tensor}
\end{equation}
Actually, $T$ is the result of a very specific quadratic combination of
irreducible representations of the torsion tensor under the Lorentz group $%
SO(1,3)$, see \cite{Hehl}. Equation \eqref{divergence} simply says that the
Weitzenb\"{o}ck invariant $T$ differs from the scalar curvature $R$ in a total
derivative; therefore, both conceptual frameworks are totally equivalent at
the time of describing the dynamics of the gravitational field. This also implies that $T$ is the unique combination of quadratic torsion terms which is locally Lorentz invariant up to a surface term. 

In many ways $f(T)$ gravity can be viewed as a natural extension of Einstein gravity in its teleparallel form, similar to $f(R)$ gravity in the more standard metric formulation. It is governed by the following action in $D$ spacetime dimensions
\begin{equation}
    S=\frac{1}{16 \pi G}\int \left[ f(T) +L_{\rm matter}\right] e\, d^{D}\negmedspace x\,.
    \label{actionfT3D0}
\end{equation}
Of course GR is contained in (\ref{actionfT3D0}) as the particular case when $f(T)=T$. The dynamical equations of $f(T)$ gravity theories are obtained after varying the action (\ref{actionfT3D0}) with respect to the vielbein components. The matter fields couple to the metric in the usual way so that the field equations become
\begin{equation}
 \left(e^{-1}\ \partial_\mu(e\ S_a^{\ \ \mu\nu})\, +\, e_a^\lambda \
T^\rho_{\ \ \mu\lambda}\ S_\rho^{\ \ \mu\nu}\right)\, f^{\prime }(T)\, + 
 S_a^{\ \ \mu\nu}\ \partial_\mu(T)\ f^{\prime \prime }(T)\, -\frac{1}{4}
e_a^\nu\ f(T)\ =\ -4\pi G\ e_a^\lambda\ T_\lambda^{\ \nu}\ , \label{ecuaciones}
\end{equation}
where the prime means derivative with respect to $T$ and $T_\lambda^{\ \nu}$ is the energy-momentum tensor.

As pointed out several times (see \cite{Nostodos} for a recent discussion), the equations (\ref{ecuaciones}) in the general case when $f''(T) \neq 0$ are sensitive to the local orientation of the vielbein. This is because they determine the entire set of components $e_a^\mu(x)$, and not just the subset related to the metric tensor $g^{\mu\nu}=e_a^\mu e_b^\nu \eta^{ab}$. In other words, Eqs. (\ref{ecuaciones}) define the spacetime structure by means of a parallelization by $D$ non-null and smooth vector fields $e_a^\mu(x)$. It is natural, then, that the \emph{vielbein grid} so determined is sensitive to local boosts and rotations $\Lambda_{a}{}^{b}(x)$ acting on it according to $e_a^\mu\rightarrow e_{b}^\mu=\Lambda_{b}{}^{a}\,e_a^\mu$. These local Lorentz transformations $\Lambda_{a}{}^{b}(x)$ have the effect of breaking the global structure of $e_a^\mu(x)$, turning the vielbein grid into an uncorrelated set of orthonormal bases at different points of the tangent space $\mathcal{T}_p(\mathcal{M})$, however leaving the metric invariant. It is important to mention that the breaking of the Lorentz covariance is relevant at scales where $f''(T)$ is considerably different from zero, for otherwise $f(T)\rightarrow T$ and the full Lorentz group is restored through TEGR. Scales where $|f''(T)|\gg 1$ are relevant concerning the manifestation of new degrees of freedom \cite{RyM,RyM2}.

On purely mathematical terms, there exist equivalence classes of vielbein grids, because parallelizations (if they exist) are non-unique. In the context of $f(T)$ theories this symmetry is realized by means of the \emph{remnant group} of Lorentz transformations of a given spacetime \cite{Nos2015}, about which we shall briefly comment in Appendix \ref{remnantgroup}. On an operational level, in turn, the restricted (remnant) group usually is not enough to impose symmetries on the vielbein if one only knows the symmetries of the metric tensor; this makes it complicated to anticipate the structure of the fields $e_a^\mu(x)$ by having only certain knowledge on the structure of $g_{\mu\nu}$, for infinitely many $e_a^\mu(x)$'s correspond to the same metric $g_{\mu\nu}$. In GR (or TEGR), the infinite set of $e_a^\mu(x)$ is not a problematic issue because every pair of vielbeins is connected by a local Lorentz transformations which are part of the symmetry of the theory. In stark contrast, within $f(T)$ gravity, it is precisely the local orientation of the vielbein which becomes important. A simple discussion on the importance of parallelizability in $f(T)$ gravity can be consulted in Ref. \cite{Nostodos}. We are now ready to construct suitable fields $e_a^\mu(x)$ for a number of different cosmological manifolds in eleven dimensions.

\section{Extra dimensions given by the topological product of parallelizable spheres} \label{sec:spherical}

\subsection{General considerations}

Let us now discuss the general structure of the vielbein field when the $D$ dimensional manifold is given by the topological product of parallelizable submanifolds. This is a very subtle point in view of the lack of local Lorentz invariance of $f(T)$ theories of gravity, a point which is particularly relevant in the strong field regime. In this section we will obtain the proper parallel one-form fields of the manifolds in consideration, postponing the discussion regarding their uniqueness until Appendix \ref{remnantgroup}. With the frame fields so obtained, we will proceed to compute the field equations for the specific cases in which the seven extra dimensions are given by topological products of parallelizable spheres. 

In what follows, we are interested in a cosmological setting where the four-dimensional space is described by the flat FLRW manifold with local pseudo Euclidean coordinates $(t,x,y,z)=(t,x_{n})$ with $n=1,2,3$. The manifold structure in $D$ dimensions is chosen to be
\begin{equation}
    \mathcal{M}_{D}=\mathcal{M}_{\rm FLRW} \times \mathcal{M}_{in},
\end{equation}
where $\mathcal{M}_{\rm FLRW} = R^{4}$ with a frame $E(R^{4})$ whose components are given by
\begin{equation}
    \label{tetbulk}
    E^{t}(R^{4})=dt \,, \qquad
    E^{n}(R^{4})=a_{0}(t)\,dx_{n} \,,
\end{equation}
leading to the line element 
\begin{equation}
    \label{metbulkgen}
    ds^{2}_{\rm FLRW} = dt^{2} - a_{0}^{2}(t)(dx^{2}+dy^{2}+dz^{2})\,.
\end{equation}

It is worth mentioning that the frame defined in (\ref{tetbulk}) is by no means simply a choice. Even though many other proper tetrads exist for the description of $\mathcal{M}_{\rm FLRW}$, related to (\ref{tetbulk}) through transformations of the remnant group, the autoparallel curves of flat Euclidean space are given by straight lines which can be generated by the coordinate basis $\partial_{x_{n}}$, whose dual co-basis is $dx_{n}$. The Euclidean grid so obtained is unaltered by the conformal scale factor, which only depends on time, turning (\ref{tetbulk}) into the simplest and most transparent vierbein field describing $\mathcal{M}_{\rm FLRW}$.

Next, it will be assumed that the topological structure of $\mathcal{M}_{in}$ is
\begin{equation}
    \mathcal{M}_{in} = S^{j_{1}}\times S^{j_{2}}\times\ldots\times S^{j_{k}} \,,
    \qquad \sum_{k} j_{k}=D-4 \,,
\end{equation}
where $S^{j_{k}}$ is the $j_{k}-sphere$. In general $\mathcal{M}_{in}$ is not a parallelizable manifold, but it turns out to be if at least one of the $j_{k}$ is odd \cite{Kervaire} and, if this would be the case, it was shown that explicit parallelizations can be found \cite{Bruni,Parton}. However, it is clear that even if (at least) one the $j_{k}$ is odd, the structure of the one-forms associated to a parallelization of $\mathcal{M}_{in}$, does not inherit the product structure of the space, because just three spheres are parallelizable by themselves, recall that these are $S^{1}$, $S^{3}$, $S^{7}$, see \cite{Ker}. If we fix $D=11$ and we work only with the three parallelizable spheres, we have four possible structures concerning the parallel one-form fields of $\mathcal{M}_{in}$, namely

\begin{eqnarray}
    \label{cuatpos}
    E_{1}(\mathcal{M}_{in})&=&E(T^{7}),\\ \notag
    E_{2}(\mathcal{M}_{in})&=&E(T^{4})\times E(S^{3}),\\ \notag
    E_{3}(\mathcal{M}_{in})&=&E(S^{1})\times E(S^{3})\times E(S^{3}),\\ \notag
    E_{4}(\mathcal{M}_{in})&=&E(S^{7}), 
\end{eqnarray} 
where $T^{j}=S^{1}\times...\times S^{1}$ is the $j-$torus. In this way, the full spacetime vielbein will have the structure 

\begin{equation}  \label{esquema7D}
E^{a}_{\mu}=\left(
\begin{array}{ccccccc}
1\,& 0 & 0 &\,\,\,\, 0\,\,\,\,\,\,\,\vline &  &  &   \,\,\,\,\, \\ 
0\,& a_{0}(t) & 0 & \,\,\,\,0\,\,\,\,\,\,\,\vline &  & \mathbb{O} & \\
0\,& 0 & a_{0}(t)&\,\,\, \,0 \,\,\,\,\,\,\,\vline &  &  & \\
0\,& 0 & 0 & a_{0}(t) \,\vline &  &  &  \\ \hline
&  &  & \,\,\,\,\,\,\,\,\,\,\,\,\,\,\vline & & & \\
&  & \mathbb{O} & \,\,\,\,\,\,\,\,\,\,\,\,\,\,\vline &  & E_{i}(\mathcal{M}_{in})& \\
&  &  & \,\,\,\,\,\,\,\,\,\,\,\,\,\,\vline &  &  & %
\end{array}
\right),
\end{equation}
where $E_{i}(\mathcal{M}_{in})$ formally refers to any of the fields (\ref{cuatpos}). In turn, the full space-time metric will be given by
\begin{equation}
    \label{metgendig11}
    ds^{2}=dt^{2}-a_{0}^{2}(t)(dx^{2}_{1}+dx^{2}_{2}+dx^{2}_{3})-ds^{2}_{in} \,,
\end{equation}
where $ds^{2}_{in}$ is the line element corresponding to the internal dimensions given by any of the four possible forms (\ref{cuatpos}).

In order to apply these models to a cosmological setting, we will assume a perfect fluid with energy density $\rho(t)$ and pressure $p(t)$ as the only matter source in the field equations. This means we have
\begin{equation}
    T^{\mu\nu}=(\rho+p)V^{\mu}V^{\nu}+p \,g^{\mu\nu},
    \label{tmunu}
\end{equation}
where $V^{\mu}$ is the tangent vector to the congruence of curves defining the stream lines of the fluid. In the comoving frame of the fluid, the energy-momentum tensor takes the simple form
\begin{equation}
    T^{\mu}_{\nu}=\mathrm{diag}(\rho,-p_{0},-p_{0},-p_{0},-p_{1},\ldots,-p_{D-4})\,.
    \label{TeneDos}
\end{equation}
We proceed now to characterize the global one-form fields of any of the internal manifolds mentioned in (\ref{cuatpos}), and to obtain the field equations which are implied by them.  

\subsection{$\mathcal{M}_{in}=T^{7}$}

Let us begin with probably the simplest case, since $T^{7}=S^{1} \times\ldots\times S^{1}$ is just the topological product of trivially parallelizable manifolds. In the previous article \cite{Fiorini:2013hva} we have analyzed the structure of the vielbein field and the relevant equations of motion for the general case given by $T^{j}$, so we shall revisit the main results here and focus on $T^7$. If we consider coordinates $X_{j}$ on $S^{j}$, it becomes trivial to parallelize the full spacetime $T^{7}$ by means of the vielbein field (no summation in $j$) 
\begin{equation}
    E^j(T^{7})=a_{j}(t)\,dX_{j}\,,
    \qquad 1\leq j \leq7 \,,
    \label{vielD}
\end{equation} 
where $a_{j}$ is the time dependent scale factor corresponding to each of the spheres $S^{j}$. The Weitzenb\"{o}ck invariant $T$ associated to the entire manifold $\mathcal{M}_{11}=R^{4} \times T^{7}$ (by means of (\ref{tetbulk}) and (\ref{vielD})) is given by
\begin{equation}
    T=-6\left( H_{0}^{2}+H_{0}\,\sum_{j=1}^{7}H_{j}+\frac{1}{6}
    \left[ \sum_{i,j=1}^{7} H_{i}\,H_{j} - \sum_{i=1}^{7} H_{i}^2 \right]
    \right)\,,  
    \label{inv6deA}
\end{equation}
where we used $H_{0}=\dot{a_{0}}/a_{0}$ and $H_{j}=\dot{a_{j}}/a_{j}$ for the corresponding Hubble functions. The dynamics of the various scale factors is determined by the Eqs. (\ref{ecuaciones}) which in the present case give the Hubble constraint equation
\begin{equation}
    f-2 \,T f^{\prime }=16 \pi G \rho \,.  
    \label{eqvi6d}
\end{equation}
Next, we have the three (identical) spatial equations coming from the standard FLRW part
\begin{equation}
    \label{eqesp1}
    2f' \left(  (2H_0 + \sum_{n=1}^{7} H_n)^2+H_0 (2H_0 + \sum_{n=1}^{7} H_n) +2\dot{H}_0 + \sum_{n=1}^{7} \dot{H}_n \right) 
    + 2 f'' \dot{T} (2H_0 + \sum_{n=1}^{7} H_n) +f = -16 \pi G p_0 \,.
\end{equation}
Finally there are seven additional equations which are
\begin{multline}
    \label{eqesp2}
  2f' \left( (3H_0+\sum_{n=1, n\neq j}^{7} H_n)^2+H_j (3H_0+\sum_{n=1, n\neq j}^{7} H_n)+ 3 \dot{H}_0+\sum_{n=1, n\neq j}^{7} \dot{H}_n  \right) \\
  +2 f'' \dot{T} \left( 3H_0+\sum_{n=1, n\neq j}^{7} H_n \right) +f = -16 \pi G p_j \,.
\end{multline}
As in standard cosmology, equation (\ref{eqvi6d}) coming from the temporal coordinate is the constraint, which plays the role of a modified Friedmann equation. Note that (\ref{eqesp2}) contains seven different equations which  correspond to the seven different pressures $p_{j}$ appearing in $T^{\mu}_{\nu}$. There is no \emph{a priori} reason to assume these different pressures to be the same. This simple observation motivates the study of spaces made up of products of higher dimensional spheres which introduces few scales factor, in the case of $S^7$ one will only introduce one additional scale factor.

\subsection{$\mathcal{M}_{in}=T^{4}\times S^{3}$}

Next we consider the case where the parallel one-forms have the topological structure $E(\mathcal{M}_{in})=E(T^{4})\times E(S^{3})$. If the coordinates on $T^{4}$ are $X_{j}$, we simply have, as in (\ref{vielD}), that $E^{j}(T^{4})=a_{j}(t) dX^{j}$, with $1\leq j\leq4$. This means we must now focus on the 3-sphere part of the geometry $E(S^{3})$. The fact that $S^{3}$ has a maximum number of global, non-null vector fields, is a consequence of the fact that any three-dimensional orientable manifold is parallelizable \cite{Sti}. An explicit parallelization is, however, not that trivial to find. The usual way is to view $S^{3}$ as the unit quaternions, and then, to realize that it has a non-abelian Lie group structure induced by quaternion multiplication, which in turn, induces a right translation on $S^{3}$. In this way a parallelization can be obtained by applying the right translation to a basis of vectors at the unit element of the group. The associated one-forms fields are obtained by means of the standard inner product on $R^{4}$. If coordinates $X_{j}$ $5\leq j\leq8$ are set up in $R^{4}$, a globally defined basis on $\mathcal{T}^{\ast}(S^{3})$ (up to a time dependent conformal factor) is  
\begin{eqnarray} 
    \label{camposens3}
    E^{5}(S^{3})&=&X_{6}dX_{5}-X_{5}dX_{6}-X_{8}dX_{7}+X_{7}dX_{8}\,,\\ 
    \notag
    E^{6}(S^{3})&=&X_{8}dX_{5}-X_{7}dX_{6}+X_{6}dX_{7}-X_{5}dX_{8}\,,\\ 
    \notag
    E^{7}(S^{3})&=&-X_{7}dX_{5}-X_{8}dX_{6}+X_{5}dX_{7}+X_{6}dX_{8}\,.
\end{eqnarray}
After changing coordinates by means of (\ref{coordinat}), we can write the line element of the internal dimensions as 
\begin{equation}
    ds^2_{in} =\sum_{j=1}^{4}a_{j}^{2}(t)dX_{j}^{2}+a_{5}^{2}(t)\,d\Omega_{3}^{\,2} \,,
\end{equation}
where $d\Omega_{3}^{\,2}$ is the line element of the three-sphere with coordinates $(\theta_{1},\theta_{2},\phi)$
\begin{equation}
    d\Omega_3^{\,2} = d\theta_{1}^2 + \sin\negmedspace^2\theta_{1}\, d\theta_{2}^2 +\sin\negmedspace^2\theta_{1} \sin\negmedspace^2\theta_{2}\, d\phi^2\,.
\end{equation}

Using the frame fields set up in this way, we can write the torsion scalar in the following way
\begin{align}
	T = \frac{6}{a_5^2} - 6(H_0^2 + H_5^2) - 18 H_5 \sum_{i=0}^{4}H_i + 
    (12 H_5 - 4H_0)\sum_{i=1}^{4}H_i -
    2\sum_{i=0}^{4}\left(H_i\sum_{j=i+1}^4H_j\right).
\end{align}
Note that the spacetime is characterized by six different scale factors, one scale factor of the FLRW part, one scale factor for any of the four 1-spheres, and finally one associated with the 3-sphere. Consequently, field equations are rather involved and somewhat complicated. The final result of those field equations begins with the Hubble constrain equation
\begin{multline}
    \label{valint4s3}
    f+f' \Big(\frac{6}{a_5^2}+6 H_0^2+6 \left(H_1+H_2+H_3+H_4+3 H_5\right) H_0+2 \Big(3 H_5^2+3 \left(H_3+H_4\right) H_5+H_3 H_4 \\+H_2 \left(H_3+H_4+3 H_5\right)+
    H_1 \left(H_2+H_3+H_4+3 H_5\right)\Big)-T\Big) =16 \pi G \rho \,.
\end{multline}
Next comes the equation for the FLRW pressure term
\begin{multline} 
    f+f' \Bigl[\frac{6}{a_5^2}+6 H_0^2+4 \left(H_1+H_2+H_3+H_4+3 H_5\right) H_0+4 \dot{H_0}+2 \dot{H_1}+2 \dot{H_2}+2 \dot{H_3}+2 \dot{H_4}+6 \dot{H_5} \\+2 \Big(H_1^2+\big(H_2+H_3+
    H_4+3 H_5\big) H_1+H_2^2+H_3^2+H_4^2+6 H_5^2+H_3 H_4+3 \left(H_3+H_4\right) H_5 \\
    +H_2 \left(H_3+H_4+3 H_5\right)\Big)-T\Bigr] +2 \big(2 H_0+H_1+H_2+
    H_3+H_4+3 H_5\big) \dot{T} f'' = -16 \pi G p_0 \,.
\end{multline}
The next four equations can be conveniently written in the form 
\begin{multline}
    f+f' \Bigl[\frac{6}{a_5^2}+6 \dot{H_0}+2 \sum_{n=1,n\neq j}^{4} \dot{H_i}+6 \dot{H_5}+2 \Big(6 H_0^2+3\sum_{n=1, n\neq j}^{4} H_i H_0+9 H_5 H_0+\sum_{n=1,n\neq j}^{4}H_i^2+6 H_5^2+ \\
    3 \sum_{n=1,n\neq j}^{4}H_i H_5+ \sum_{m=1,m\neq j}^{3}\sum_{n=m+1,n\neq j}^{4} H_m H_n \Big)-T\Bigr]+2 \Bigl(3 H_0+\sum_{n=1,n\neq j}^{4}H_i+3 H_5\Bigr) \dot{T} f'' = -16 \pi G p_j\,,
\end{multline}
where $j=1,2,3,4$. The final equation involving the pressure $p_5$ is given by

\begin{multline} 
    \label{esp6t4s3}
    f+f' \Bigl[\frac{2}{a_5^2}+12 H_0^2+6 \left(H_1+H_2+H_3+H_4+2 H_5\right) H_0+6 \dot{H_0}+2 \dot{H_1}+2 \dot{H_2}+2 \dot{H_3}+2 \dot{H_4}+4 \dot{H_5} \\
    +2 \big(H_1^2+\big(H_2+H_3+
    H_4+2 H_5\big) H_1+H_2^2+H_3^2+H_4^2+3 H_5^2+H_3 H_4+2 \left(H_3+H_4\right) H_5\\
    +H_2 \left(H_3+H_4+2 H_5\right)\big)-T\Bigr] +2 \big(3 H_0+H_1+H_2+
    H_3+H_4+2 H_5\big) \dot{T} f'' =  -16 \pi G p_5 \,.
\end{multline}

\subsection{$\mathcal{M}_{in}=S^{1}\times S^{3}\times S^{3}$}

Once a parallelization for $S^{3}$ is obtained, as described in the previous case, the characterization of $S^{1}\times S^{3}\times S^{3}$ proceed straightforwardly. Let $X^{1}$ be a coordinate on the circle, it follows that a global basis for $\mathcal{T}^{\ast}(S^{1})$ is given by $E^{1}(S^{1})=a_{1}(t)dX^{1}$. The parallelization of the remaining six dimensional manifold $S^{3}\times S^{3}$ is obtained by means of two copies of the fields given in (\ref{camposens3}). Up to  time dependent conformal factors in the corresponding 3-spheres, we have
\begin{eqnarray} 
\label{camposens31}
    E^{2}(S^{3})&=&X_{3}dX_{2}-X_{2}dX_{3}-X_{4}dX_{3}+X_{3}dX_{4}\,,\\ 
    \notag
    E^{3}(S^{3})&=&X_{5}dX_{2}-X_{4}dX_{3}+X_{3}dX_{4}-X_{2}dX_{5}\,,\\ 
    \notag
    E^{4}(S^{3})&=&-X_{4}dX_{2}-X_{5}dX_{3}+X_{2}dX_{4}+X_{3}dX_{5}\,,
\end{eqnarray}
for the first 3-sphere and accordinly for the second one
\begin{eqnarray} 
    \label{camposens32}
    E^{5}(S^{3})&=&X_{6}dX_{5}-X_{5}dX_{6}-X_{8}dX_{7}+X_{7}dX_{8}\,,\\
    \notag
    E^{6}(S^{3})&=&X_{8}dX_{5}-X_{7}dX_{6}+X_{6}dX_{7}-X_{5}dX_{8}\,,\\ 
    \notag
    E^{7}(S^{3})&=&-X_{7}dX_{5}-X_{8}dX_{6}+X_{5}dX_{7}+X_{6}dX_{8}\,.
\end{eqnarray}
One must take note of the common coordinate (here $X_{5}$). This is necessary in order to embed the six-dimensional manifold in $R^{7}$. The internal metric in coordinates $(X_{1},\theta_{1},\theta_{2},\phi_{1},\theta_{3},\theta_{4},\phi_{2})$ give
\begin{equation}
     ds^2_{in} =a_{1}^{2}(t)dX_{1}^{2}+a_{2}^{2}(t)\,
     d\Omega_{3\,,(1)}^{\,2}+a_{3}^{2}(t)\,d\Omega_{3\,,(2)}^{\,2} \,,
\end{equation}
where we used the notation
\begin{eqnarray}
    d\Omega_{3\,,(1)}^{\,2}&=&d\theta_{1}^2 + \sin\negmedspace^2\theta_{1}\, d\theta_{2}^2 +\sin\negmedspace^2\theta_{1} \sin\negmedspace^2\theta_{2}\, d\phi_{1}^2\,, \\
    d\Omega_{3\,,(2)}^{\,2}&=&d\theta_{3}^2 + \sin\negmedspace^2\theta_{3}\, d\theta_{4}^2 +\sin\negmedspace^2\theta_{3} \sin\negmedspace^2\theta_{4}\, d\phi_{2}^2\,.
\end{eqnarray}
With the fields so obtained, we can compute the Weitzenb\"{o}ck invariant
\begin{equation}
    T=-6\Bigl(H_{0}^{2}+H_{2}^{2}+H_{3}^{2}-a_{2}^{-2}-a_{3}^{-2}+H_{1}(H_{0}+H_{2}+H_{3})+3(H_{0}H_{2}+H_{0}H_{3}+H_{2}H_{3})\Bigr).
\end{equation}
The resulting field equations in this case begin with the Hubble constraint equation
\begin{equation}
    \label{valincsis3s3}
    f+2f'(6 a_{2}^{-2}+6 a_{3}^{-2}-T) = 16 \pi G \rho \,.
\end{equation}
This is followed by the four evolution equations
\begin{eqnarray}
    \notag
    && f+2 f' \left(2 \dot{H_0}+\dot{H_1}+3 \dot{H_2}+3 \dot{H_3}+\left(2 H_0+H_1+3 \left(H_2+H_3\right)\right) \left(3 H_0+H_1+3 \left(H_2+H_3\right)\right)\right) \\
    && \qquad \qquad 
    + 2f'' \dot{T}\left(2 H_0+H_1+3 \left(H_2+H_3\right)\right)=-16 \pi G p_0\,,  \\[1ex]
    \notag
    && f+6f' \left(\dot{H_0}+\dot{H_2}+\dot{H_3}+\left(H_0+H_2+H_3\right) \left(3 H_0+H_1+3 \left(H_2+H_3\right)\right)\right) \\
    && \qquad \qquad
    + 6 f''\dot{T} \left(H_0+H_2+H_3\right)=-16 \pi G p_1\,, \\[1ex]
    && f+2f' \left(3 \dot{H_0}+\dot{H_1}+2 \dot{H_2}+3 \dot{H_3}+\left(3 H_0+H_1+2 H_2+3 H_3\right) \left(3 H_0+H_1+3 \left(H_2+H_3\right)\right)-2 a_2^{-2}\right)\\ 
    \notag
    && \qquad \qquad
    + 2f'' \dot{T} \left(3 H_0+H_1+2 H_2+3 H_3\right) =-16 \pi G p_2\,, \\[1ex]
    \notag
    && f+2 f'\left(3 \dot{H_0}+\dot{H_1}+3 \dot{H_2}+2 \dot{H_3}+\left(3 H_0+H_1+3 H_2+2 H_3\right) \left(3 H_0+H_1+3 \left(H_2+H_3\right)\right)-2a_3^{-2}\right)\\ 
    && \qquad \qquad
    + 2 f'' \dot{T} \left(3 H_0+H_1+3 H_2+2 H_3\right) =-16 \pi G p_3\,.
    \label{espcsis3s3}
\end{eqnarray}
The entire set of field equations can be viewed as four dynamical equations for the four Hubble functions $H_i$, $i=0,1,2,3$ subject to the constraint (\ref{valincsis3s3}). For given $f(T)$ one could attempt a dynamical systems formulation in order to understand the dynamics of such a cosmological model, this is what will be done with the final case which is studied next.

\subsection{$\mathcal{M}_{in}=S^{7}$}

It has been known for a long time that $S^{7}$ is parallelizable, it is nonetheless surprising that explicit expressions for global bases of vector fields in $\mathcal{T}(S^{7})$ (or one-forms in $\mathcal{T}^{\ast}(S^{7})$), are rarely found in the literature (see, e.g. \cite{Irina}). The procedure in order to obtain a parallelization is to view $S^{7}$ as the unit octonion, and to use their multiplication rule to obtain right invariant vector fields. Due to the fact that multiplication of unit octonions is not associative, $S^{7}$ is not a Lie group, however this is not an impediment in getting a global basis of vector fields on $S^{7}$. Let us choose coordinates $X_{i}$, $i=1,\ldots,8$ in $R^{8}$, a global basis of one-forms in $\mathcal{T}^{\ast}(S^{7})$ can be written explicitly (up to a time dependent conformal factor) as follows
\begin{eqnarray} 
    \label{camposens32a}
    E^{1}(S^{7})&=&-X_{2}dX_{1}+X_{1}dX_{2}-X_{4}dX_{3}+X_{3}dX_{4}-X_{6}dX_{5}+X_{5}dX_{6}-X_{8}dX_{7}+X_{7}dX_{8}\,,\\ 
    \notag
    E^{2}(S^{7})&=&-X_{3}dX_{1}+X_{4}dX_{2}+X_{1}dX_{3}-X_{2}dX_{4}-X_{7}dX_{5}+X_{8}dX_{6}+X_{5}dX_{7}-X_{6}dX_{8}\,,\\ 
    \notag
    E^{3}(S^{7})&=&-X_{4}dX_{1}-X_{3}dX_{2}+X_{2}dX_{3}+X_{1}dX_{4}+X_{8}dX_{5}+X_{7}dX_{6}-X_{6}dX_{7}-X_{5}dX_{8}\,,\\ 
    \notag
    E^{4}(S^{7})&=&-X_{5}dX_{1}+X_{6}dX_{2}+X_{7}dX_{3}-X_{8}dX_{4}+X_{1}dX_{5}-X_{2}dX_{6}-X_{3}dX_{7}+X_{4}dX_{8}\,,\\
    \notag       
    E^{5}(S^{7})&=&-X_{6}dX_{1}-X_{5}dX_{2}-X_{8}dX_{3}-X_{7}dX_{4}+X_{2}dX_{5}+X_{1}dX_{6}+X_{4}dX_{7}+X_{3}dX_{8}\,,\\
    \notag
    E^{6}(S^{7})&=&-X_{7}dX_{1}+X_{8}dX_{2}-X_{5}dX_{3}+X_{6}dX_{4}+X_{3}dX_{5}-X_{4}dX_{6}+X_{1}dX_{7}-X_{2}dX_{8}\,,\\
    \notag 
    E^{7}(S^{7})&=&-X_{8}dX_{1}-X_{7}dX_{2}+X_{6}dX_{3}+X_{5}dX_{4}-X_{4}dX_{5}-X_{3}dX_{6}+X_{2}dX_{7}+X_{1}dX_{8}\,.
\end{eqnarray}
After changing to hyperspherical coordinates $(\theta_{1},...,\theta_{6},\phi)$ the line element in $S^{7}$ becomes
\begin{equation}
  ds^2_{in} = a_{1}^{2}(t)\,d\Omega_7^{\,2} \,,
\end{equation}
where the line element of the 7-sphere is
\begin{equation}
    d\Omega_7^{\,2} = d\theta_{1}^2+\sin\negmedspace^2\theta_{1}\,d\theta_{2}^2+...+d\phi^2\prod_{i=1}^{6}\sin\negmedspace^2\theta_{i} \,.
\end{equation}
The torsion scalar $T$ reads 
\begin{equation}
    \label{invariant}
    T=-6(H_{0}^{2}+7H_{0}H_{1}+7H_{1}^{2}-7a_{1}^{-2}) \,.
\end{equation}
Finally we can state the complete set of field equations. Due to the appearance of only one additional scale factor in this model, we expect somewhat simpler equations than in the previous case. This turns out to be the case as can be seen in the following. The temporal field equation simply becomes
\begin{equation}
    \label{primera}
    f+2f'(42a_{1}^{-2}-T)=16\pi G \rho \,,
\end{equation}
while the two dynamical equations are given by
\begin{eqnarray}
    f+2f'\Big(6H_{0}^{2}+35H_{0}H_{1}+49H_{1}^{2}+2\dot{H}_{0}+7\dot{H}_{1}\Big)+
    2f'' \dot{T}\Big(2H_{0}+7H_{1}\Big) &=& -16\pi G p_{0}\,,
    \label{ecespa} \\
    f+6f'\Big(3H_{0}^{2}+13H_{0}H_{1}+14H_{1}^{2}-2a_{1}^{-2}+\dot{H}_{0}+2\dot{H}_{1}\Big)+
    3f''\dot{T}\Big(H_{0}+2H_{1}\Big) &=& -16\pi G p_{1}\,.
    \label{ecespc}
\end{eqnarray}
Compared with the previously discussed cases where the extra dimensions contain various products of spheres, the field equation for $S^7$ are much simpler to deal. In what follows we will discuss those equations in some detail and show that they contain many desirable features when considering applications to realistic cosmological models.

\section{Some consequences of the field equations}
\label{dynamic}

\subsection{Early inflation powered by extra dimensions}

Let us assume for the moment that the scale factors corresponding to the internal dimensions are constant, meaning that all Hubble functions other than $H_0$ vanish identically. Presumably, this could represent a good approximation to the final stages of the evolution where the additional dimensions no longer affect the universe, which is then governed solely by the scale factor $a_{0}(t)$ of the four dimensional, spatially flat FLRW metric. In this case, it is not hard to see that the full system of equations (\ref{ecuaciones}) decouples into two sets of equations which are very different in structure. One of the sets correspond to the usual $f(T)$ equations associated with a four dimensional, $K=0$, FLRW cosmological model. The equations within this set completely determine the scale factor $a_{0}(t)$ of the physical macroscopic dimensions. The other set of equations, on the other hand, consists of algebraic equations relating the pressures of the internal dimensions. Let us henceforth look into this in some detail.

If we start with the $S^{7}$ case, the imposition of constant $a_{1}$ lead us to torsion invariant (\ref{invariant}) which together with its time derivative read
\begin{equation}
    \label{torandder}
    T=-6(H_{0}^{2}-7a_{1}^{-2})\,,\qquad
    \dot{T}=-12H_{0}\dot{H}_{0} \,.
\end{equation}
In turn, the equations of motion (\ref{primera})--(\ref{ecespc}) can be written as
\begin{eqnarray}
    \label{primeracons}
    f+12f'H_{0}^{2} &=& 16\pi G \rho\,, \\
    f+4f'\bigl(3H_{0}^{2}+\dot{H}_{0}\bigr)-48f''H_{0}^{2}\dot{H}_{0} &=& -16\pi G p_{0}\,,\label{ecespaconst} \\
    f+6f'\bigl(3H_{0}^{2}+\dot{H}_{0}-2a_{1}^{-2}\bigr)-36f''H_{0}^{2}\dot{H}_{0} &=& -16\pi G p_{1}\,.
    \label{ecespcconst}
\end{eqnarray}

Eqs. (\ref{primeracons}) and (\ref{ecespaconst}) have the same structure as the cosmological field equations of a spatially flat FLRW cosmology in four-dimensional $f(T)$ gravity \cite{Bengochea:2008gz}. The only difference comes from the Weitzenbock scalar which includes the constant scale factor of the extra dimensions. Hereafter, we will refer to Eqs. (\ref{primeracons}) and (\ref{ecespaconst}) as the $4D$ \emph{reduced} $f(T)$ \emph{equations}.

The role of the constant $a_{1}$ in the expression for $T$ can be easily appreciated if we consider the GR case. Taking $f=T$, $f'=1$ and $f''=0$ in the system (\ref{primeracons})--(\ref{ecespcconst}) we arrive at
\begin{eqnarray}
    \label{primeraconsgr}
    H_{0}^{2}+7a_{1}^{-2} &=& \frac{8}{3}\pi G \rho \,, \\
    H_{0}^{2}+\frac{2}{3}\dot{H}_{0}+7a_{1}^{-2} &=& -\frac{8}{3}\pi G p_{0}\,,
    \label{ecespaconstgr} \\
    2H_{0}^{2}+\dot{H}_{0}+5a_{1}^{-2} &=& -\frac{8}{3}\pi G p_{1}\,.
    \label{ecespcconstgr}
\end{eqnarray}
Despite the absence of the cosmological constant in the original action, the effect of the constant scale factor of the extra dimensions is to generate a negative `effective' cosmological constant given by $\Lambda = -21a_{1}^{-2}$. The sign of $\Lambda$ is fixed to be negative which comes from the expression of the Weitzenbock scalar (\ref{torandder}). There $H_{0}^{2}$ and $a_{1}^{-2}$ enter with different signs.

In the general setting where the function $f(T)$ is arbitrary, we can solve equations (\ref{primeracons}) and (\ref{ecespaconst}) for a given matter source content, and then obtain the pressure of the internal space by means of (\ref{ecespcconst}). Due to the automatic conservation of $T_{\mu\nu}$ in the external space, the two equations (\ref{primeracons}) and (\ref{ecespaconst}) are not independent. Consequently it is enough to solve (\ref{primeracons}) with $\rho(t)$ given by $\rho(t)=\rho_{0}(a_{0}(t))^{-3(1+\omega)}$. This factorization is possible primarily because $a_{1}$ is assumed to be constant, and therefore it is absent in the conservation equation of the fluid. In general, we have
\begin{equation}
    \dot{\rho}+(3 H_{0}+\sum_{n=1}^{D-4}H_{n})\rho +3H_{0}p_{0}+ \sum_{n=1}^{D-4}H_{n}p_{n}=0 \,,
    \label{formacons}
\end{equation}
which shows, in the present situation, that the pressure $p_{1}$ obtained by means of Eq. (\ref{ecespcconst}), has no effect whatsoever on the energy density of the 4-dimensional Universe we experience. This is so because $H_{1}$ vanishes.

The above discussion extends nicely to the other topologies we considered. In the $T^7$ case, Eqs. (\ref{eqvi6d})--(\ref{eqesp2}) for constant scale factors $a_{i}$ $i=1,\ldots,7$ become the $4D$ reduced $f(T)$ equations, together with the one additional equation
\begin{equation}
    f+6f'\bigl(3H_{0}^{2}+\dot{H}_{0}\bigr)-36f''H_{0}^{2}\dot{H}_{0} = -16\pi G p_{j}\,,\qquad j=1,\ldots,7 \,.
    \label{ecespaconst7-toro}
\end{equation} 
In case of topology $T^4 \times S^3$ the corresponding algebraic relations are 
\begin{eqnarray}
    f+6f'\bigl(3H_{0}^{2}+\dot{H}_{0}\bigr)-72f''H_{0}^{2}\dot{H}_{0} &=& -16\pi G p_{j}\,,\qquad j=1,\ldots,4 \,.
    \label{ecespcconst4}\\
    f+2f'\bigl(9H_{0}^{2}+3\dot{H}_{0}-2a_{5}^{-2}\bigr)-72f''H_{0}^{2}\dot{H}_{0} &=& -16\pi G p_{5} \,.
    \label{ecesp2cconst4}
\end{eqnarray}

The case $S^1 \times S^3\times S^3$ follows the similar lines. Eqs. (\ref{valincsis3s3})--(\ref{espcsis3s3}) show that, assuming the internal scale factors $a_{1}$, $a_{2}$ and $a_{3}$ to be constants, the system leads to the $4D$ reduced $f(T)$ equations plus
\begin{eqnarray}
    f+6f'\bigl(3H_{0}^{2}+\dot{H}_{0}\bigr)-72f''H_{0}^{2}\dot{H}_{0} &=& -16\pi G p_{1} \,,
    \label{ecespcconss3s1}\\
    f+2f'\bigl(9H_{0}^{2}+3\dot{H}_{0}-2a_{j}^{-2}\bigr)-72f''H_{0}^{2}\dot{H}_{0} &=& -16\pi G p_{j}\,, \qquad j=2,3 \,.
    \label{ecesp2ccons3s1}
\end{eqnarray}

It is interesting to note that, in absence of any matter, $f(T)$ gravity is sometimes able to describe an early time de Sitter accelerated stage for the macroscopic 4-spacetime which is caused by the presence of extra dimensions. As is clear from Eqs. (\ref{primeraconsgr})--(\ref{ecespcconstgr}), in GR this cannot be achieved. Eqs. (\ref{primeraconsgr}) and (\ref{ecespaconstgr}) give $H_{0}^2=-7a_{1}^{-2}$, which is not only non-physical, but also inconsistent with (\ref{ecespcconstgr}).

In vacuum, Eqs. (\ref{primeracons})--(\ref{ecespcconst}) considering the $S^7$ case with constant $H_{0}$, reduce to the simple equations
\begin{equation}
    f+12f'H_{0}^2=0\,, \qquad 
    f+6f'(3H_{0}^2-2a_{1}^{-2})=0 \,.
    \label{ecsenvacios7}
\end{equation}
Combining them we get $a_{1}^{-2}=H_{0}^2/2$, which is valid for any $f(T)$ model other than GR. However, the value of $H_{0}$ (or $a_{1}$) depends on the function $f(T)$. For instance, let us consider an ultraviolet deformation of the form $f(T)=T+\alpha T^{2}$. Due to the fact that $a_{1}^{-2}=H_{0}^2/2$ we have that $T=15H_{0}^2=30a_{1}^{-2}$ (see Eq. (\ref{torandder})). This means that Eqs. (\ref{ecsenvacios7}) relate the constant $\alpha$ to the inflationary Hubble rate by means of 
\begin{equation}
    \label{detalfa}
    \alpha = -\frac{3}{65 H_{0}^2} = -3 \frac{a_{1}^{2}}{130} \,.
\end{equation}
Consequently, $\alpha$ should be negative and very small, by virtue of the fact that $H_{0}$ is large during inflation. This simple model enables to link the typical deformation scale $\alpha$ to the (squared) length scale $a_{1}^{2}$ characterizing the size of the extra dimensions during the inflationary era. In other words small extra dimensions give rise to large inflation.

However, not every topology enables us to describe extra dimensions-powered inflation. Note that the $4D$ reduced $f(T)$ equations (\ref{primeracons}) and (\ref{ecespaconst}) coalesce to the single equation $f+12f'H_{0}^2=0$, in vacuum and for constant $H_{0}$. Then, the $T^7$ topology leads to the inconsistent (for a non null $H_{0}$) system
\begin{equation}
    f+12f'H_{0}^2=0\,,\qquad
    f+18f'H_{0}^2=0\,.
    \label{ecsenvaciostoro7}
\end{equation} 
Similarly the $T^4 \times S^3$ case shows the same inconsistency by means of the equations
\begin{equation}
    f+12f'H_{0}^2=0\,,\qquad 
    f+18f'H_{0}^2=0\,,\qquad
    f+2f'(9H_{0}^2-2a_{5}^{-2})=0 \,.
    \label{ecsenvaciost4s3}
\end{equation}
Exactly the same happens when $S^1 \times S^3\times S^3$ is considered. The vacuum field equations turn into
\begin{equation}
    f+12f'H_{0}^2=0\,,\qquad
    f+18f'H_{0}^2=0\,,\qquad
    f+2f'(9H_{0}^2-2a_{j}^{-2})=0 \,,
    \label{ecsenvacioss1s3s3}
\end{equation}
with $j=2,3$. These results seems to suggest that the 7-sphere $S^7$ is clearly favored on physical grounds, at least, regarding the interpretation of the early inflationary era as an effect produced by the presence of the extra dimensions. This motivates us to have a closer look at the dynamical equations of that case.

\subsection{Dynamical systems analysis}

Let us now have a closer look at the equations of motion for the aforementioned model $f(T)=T+\alpha T^{2}$ which appears to have some desirable properties regarding the inflationary era. As in the above discussion, we consider the vacuum equations, this means setting $\rho=p_a=p_b=0$. Eqs.~(\ref{invariant})--(\ref{ecespc}) do contain first time derivatives of the Hubble parameters $H_0$ and $H_1$ but do not contain second derivatives. Therefore, we can, in principle, reformulate these field equations as a first order system of autonomous equations of the form
\begin{align}
    \frac{dH_0}{dt} = \mathcal{A}(H_0,H_1)\,, \qquad 
    \frac{dH_1}{dt} = \mathcal{B}(H_0,H_1)\,, 
    \label{eqdyn1}
\end{align}
where $\mathcal{A}$ and $\mathcal{B}$ are two rather complicated functions which can be stated explicitly and are given in Appendix~\ref{AppD}, see Eq.~(\ref{AppDEq1}) and~(\ref{AppDEq2}). In the following we will set $\alpha = -0.1$ since the equations are somewhat too cumbersome to deal with them generically. 

Following the standard procedure of cosmological dynamical systems, see for instance~\cite{Bahamonde:2017ize}, we begin by looking for the critical points of this systems. These are the points where the system is in equilibrium and are defined by the vanishing of the right-hand sides of~(\ref{eqdyn1}). The critical points can be found numerically and the four points and their properties are summarized in Table~\ref{tab:my_label}.

\begin{table}[!htp]
    \centering
    \begin{tabular}{|c|c|c|c|c|}
        \hline
        Point & $H_0$ value & $H_1$ value & eigenvalues $\lambda$ & properties\\
        \hline
         $A_+$ & $+0.719$ & $-0.571$ & $1.093 \pm 2.517\, i$ & unstable spiral\\
         $A_-$ & $-0.719$ & $+0.571$ & $-1.093\pm 2.517\, i$ & stable spiral\\ 
         $B_+$ & $+0.679$ & 0 & $-3.309\,;-1.157$ & stable node\\ 
         $B_-$ & $-0.679$ & 0 & $3.309\,;1.157$ & unstable node\\ 
         \hline
    \end{tabular}
    \caption{Critical points and eigenvalues of the model $f(T)=T+\alpha T^{2}$ with $\alpha=-0.1$.}
    \label{tab:my_label}
\end{table}

\begin{figure}[!htb]
\centering
\includegraphics[width=0.48\textwidth]{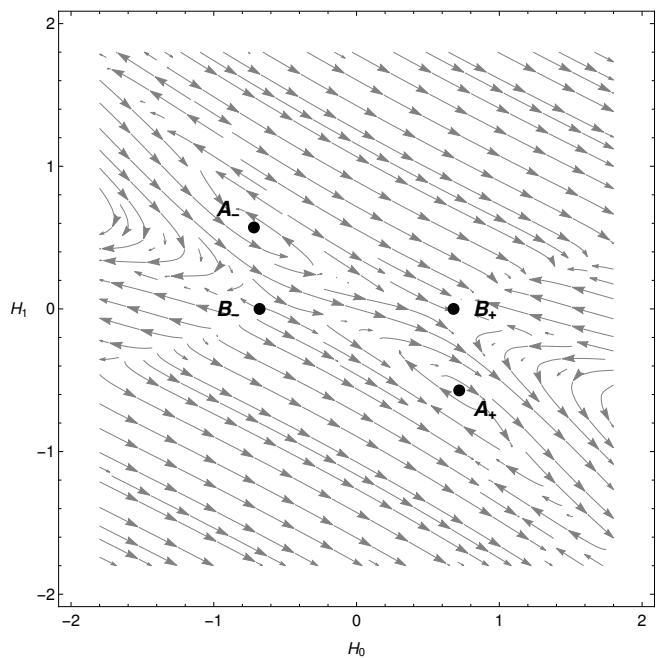}
\caption{Phase portrait near the critical points. Quadratic model $f(T)=T+\alpha T^{2}$ with $\alpha=-0.1$.}
\label{fig1}
\end{figure}

The equilibrium points $A_\pm$ in Fig.~\ref{fig1} correspond to cosmological solutions where one of the two scale factors expands while the other one contracts. The stable spiral point $A_-$ corresponds to an ever contracting universe $H_0 < 0$ in which the extra space $S^7$ expands. From a physical point of view this equilibrium point is somewhat undesirable. On the other hand, point $A_+$ describes an expanding universe $H_0 > 0$ (with contracting $S^7$) which is unstable. Such a state is suitable for an early time inflationary model as the universe would grow rapidly but then change its behavior. Recall that unstable points can be interpreted as early time attractors.

Points $B_\pm$ are characterized by the condition that the extra space becomes static $H_1 = 0$, we can either have an expanding or a contracting universe. The value of $H_{0}$ at these points is given in Eq. (\ref{detalfa}), with $\alpha = -0.1$. Interestingly, the expanding solution which corresponds to $B_+$ is a stable node. The deceleration parameter for the scale factor $a_0$ can be expressed as $q=-1-\dot{H}_0/H_0^2$ so for all critical points we obtain $q_a=-1$, since $\dot{H}_0 = 0$ by definition. Consequently all critical points are candidates for either early time or late time accelerated expansion. 

Going back to (\ref{detalfa}) using $\alpha = -1/10$ we can solve for $H_0$ and find $H_0 = \pm \sqrt{6/13}$ which corresponds to the values at the points $B_{\pm}$. Therefore, the critical point $B_{+}$ corresponds to an early time inflationary state. We notes that in this model trajectories are attracted to this state. The point $A_{+}$ would correspond to a dark energy dominated state. We also note that $A_{+}$ and $B_{-}$ are early time attractors (unstable points) and it is becoming quite clear that the phase space shows an intricate structure.

Therefore the global picture needs to be considered before making detailed conclusions. Figure~\ref{fig1} suggests the existence of various critical points at infinity and it will turn out that we cannot identify trajectories in phase space which connect points $A_+$ and $B_+$.

Before proceeding we note that the system~(\ref{eqdyn1}) is invariant under the transformation $t \mapsto -t$ and $H_{0,1} \mapsto -H_{0,1}$ which explains the observed symmetries of this system. This will becomes particularly obvious when the global phase portrait is taken into account.

\begin{figure}[!htb]
\centering
\includegraphics[width=0.48\textwidth]{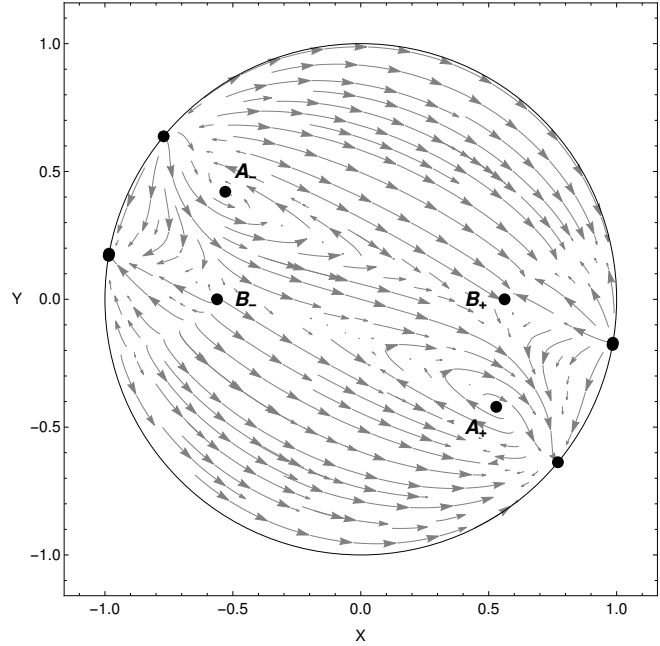}
\caption{Global phase portrait with critical points at infinity. Quadratic model $f(T)=T+\alpha T^{2}$ with $\alpha=-0.1$.}
\label{fig2}
\end{figure}

In order to show the global phase portrait of our system, we follow the standard procedure, see e.g.~\cite{Bahamonde:2017ize}, of introducing the Poincar\'e sphere thereby compactifying the entire phase space to the unit sphere. We introduce the new variables
\begin{align}
    X = \frac{H_0}{\sqrt{1+H_0^2+H_1^2}}\,, \qquad 
    Y = \frac{H_1}{\sqrt{1+H_0^2+H_1^2}}\,,
    \label{eqdyn2}
\end{align}
so that the region $H_0,H_1 \rightarrow \infty$ corresponds to the boundary of the unit circle. The critical points at infinity are found by finding the roots of the function 
\begin{align}
    G_{m+1} = X \mathcal{B}_m (X,Y) - Y \mathcal{A}_m (X, Y ) = 0 \,,
    \label{eqdyn3}
\end{align}
where $\mathcal{A}_m,\mathcal{B}_m$ stand for the highest order polynomial power in the right-hand sides of (\ref{eqdyn1}). However, this system is not polynomial in the variables so we expanded the system near infinity and extracted the leading order polynomial terms in that expansion. These are quadratic in the variables, so $m=2$, and are given by
\begin{align}
    \mathcal{A}_2 &= \frac{-15147-16894 \sin (2 \theta )+5245 \sin (4 \theta )+11920 \cos (2 \theta )+1979 \cos (4 \theta )}{8 (123-87 \cos (2 \theta )+231 \sin (\theta
   ) \cos (\theta ))} \,, 
    \label{eqdyn4} \\
    \mathcal{B}_2 &= \frac{-39345-43358 \sin (2 \theta )+14483 \sin (4 \theta )+32072 \cos (2 \theta )+4777 \cos (4 \theta )}{84 (-82-77 \sin (2 \theta )+58 \cos (2
   \theta ))} \,.
    \label{eqdyn5}
\end{align}
It turns out that these terms are independent of the parameter $\alpha$ which consequently only affects the local critical points. Given that $m=2$ the function $G_3$ of (\ref{eqdyn3}) will have at most pairs of roots. Each root $\theta_n$  ($n=1,2,3$) comes with an associated root located at $\theta_n + \pi$. The global phase portrait with critical points at infinity including the previously discussed local critical points is given in Fig.~\ref{fig2}.

It appears that there are only two pairs of critical points at infinity in Fig.~\ref{fig2}. This has to do with the fact that two of these pairs are very close to each other. The angular values $\theta_i$, $i=1,2,3$ which determine the locations of those points on the unit circle $(\cos(\theta_i),\sin(\theta_i))$ are approximately $\theta_1=2.450$, $\theta_2 = 2.962$ and $\theta_3 = 2.971$. As mentioned above, these come with their associated pair located at $\theta_i + \pi$.

\begin{figure}[!htb]
\centering
\includegraphics[width=0.48\textwidth]{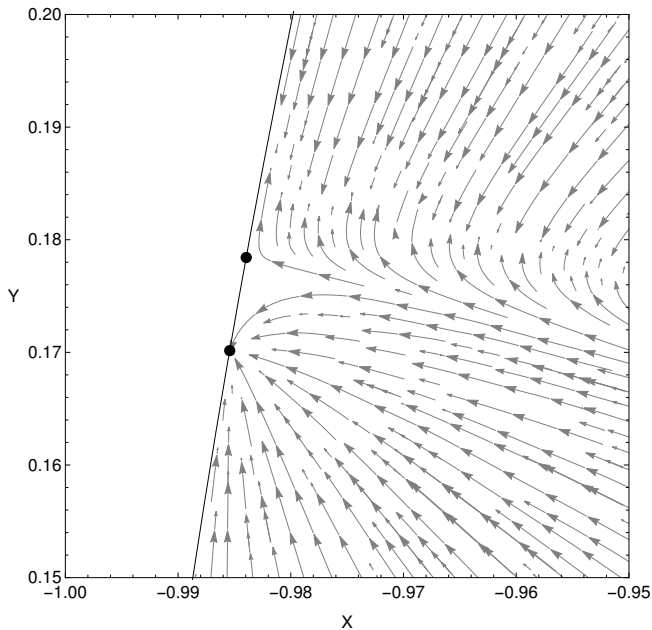}
\caption{Detailed global phase portrait near two nearby critical points at infinity. Quadratic model $f(T)=T+\alpha T^{2}$ with $\alpha=-0.1$.}
\label{fig3}
\end{figure}

This is not a numerical effect but a true feature of the dynamical system. This can be seen be showing a detailed phase portrait near these points, see Fig.~\ref{fig3}. One sees that both critical points $\theta_2$ and $\theta_3$ are `close' to each other, however, they show distinct features. The lower point attracts trajectories while the upper one repels them, a feature which is lost when the entire phase space is shown. 

From a physical point of view it is important to point out that all critical points at infinity are located in the second or fourth quadrant of the phase space. This means all solutions approaching these points will always have one contracting and one expanding Hubble parameter, in the second quadrant we have $H_0 < 0$ and $H_1 > 0$ while in the fourth quadrant we have $H_0 > 0$ and $H_1 < 0$. Only the local critical points $B_\pm$ are different in the sense that the Hubble parameters $H_1$ identically vanishes, $H_1 = 0$.

This model clearly displays a very interesting dynamical behaviour which contains various epochs where the 4-dimensional part of the manifold expands, the right half of the phase space shown in Fig.~\ref{fig2}. During the expansion of the 4-dimensional part of the manifold, the extra 7-dimensional part will eventually contract due to the location of the critical points at infinity. We can conclude that the extra dimensions, in general, affect the dynamics of the 4-dimensional part of the manifold and that inflationary epochs are naturally part of such systems. No matter sources are required to drive the expansion (or contraction) and consequently one could conclude that epochs of expansion appear naturally in such models.







\section{Concluding comments}
\label{sec:conclution}

Modified teleparallel models of gravity where studied in eleven dimensions and applications to cosmology were considered. The four-dimensional part of the manifold was assumed to be the usual FLRW manifold with flat constant time hypersurfaces while the seven extra dimensions were assumed to be products of parallelizable spheres, that is $S^{1}$, $S^{3}$ and $S^{7}$.
Using these assumptions one is led to the following four possible compactifications of the extra dimensions: $T^7$, $T^4\times S^3$, $S^1\times S^3\times S^3$, and $S^7$. For each of these cases the corresponding structure of the 1-forms field was obtained. These vielbeins constitute the starting point for any $f(T)$ cosmological model including the above mentioned compactifications, because they represent the basis responsible for the parallelization of the manifolds under consideration, i.e., they define the space-time structure. Endowed with this important information, we obtained the $f(T)$ cosmological equations for each of the cases in question.

In Section \ref{dynamic} we analyzed the structure of the equations more closely by considering the possibility of having de Sitter-like epochs in the four dimensional FLRW submanifold due to the presence of the extra dimensions. In the absence of any matter content, and fixing the Hubble factors of the extra dimensions to be vanishing, we could show that not all topologies considered give rise to situations where the extra dimensions can drive a period of accelerated expansion. In fact, for the cases analyzed, $S^7$ is clearly favored from a theoretical point of view, in particular regarding the interpretation of an early time inflationary era driven by the extra dimensions. Due to the fact that the size of the extra dimensions and the constant Hubble factor of the inflationary stage verify $a_{1}^{-2}=H_{0}^2/2$, it was shown that the dynamical equations favor the evolution towards the stable node $B_{+}$ corresponding to large exponential growth given by the smallness of the 7-sphere $S^7$.

It remains unclear why only the 7-sphere naturally leads to a vacuum inflationary stage, but it should be emphasized that this property was anticipated at the end of reference \cite{Fiorini:2013hva}. As a matter of fact, something similar happens in $D=7$, where the 3-sphere plays the role of powering vacuum inflation there, see the mentioned reference. In order to prove that $S^7$ is the sole topology driving inflation in $D=11$, we need to consider the remaining eleven different topological products of spheres possible. One of those, the case $S^3 \times S^2 \times S^2$, is considered in the appendix \ref{anotherexample}, from where it is easy to understand, again, that $S^7$ appears to be the most natural choice. Continuing along those lines, we expect to develop the appropriate techniques in order to deal with the complete set of `non-trivial' internal space parallelizations in the future.



\acknowledgments
This work was partially funded by Comisi\'{o}n
Nacional de Ciencias y Tecnolog\'{i}a through FONDECYT Grant
11140674 (PAG) and by the Direcci\'on de Investigaci\'on y Desarrollo de la Universidad de La Serena (YV).

CGB and FF acknowledge support by The Royal Society, International Exchanges 2017, grant number IEC/R2/170013. FF is member of Carrera del Investigador Científico (CONICET), and he is supported by CONICET and Instituto Balseiro. 

FF would like to thank the hospitality at Universidad Diego Portales and at the Department of Mathematics at University College London, where this investigation was conceived and partially undertaken. PAG acknowledge the hospitality of the Universidad de La Serena and Centro At\'{o}mico Bariloche where part of this work was done.

\appendix

\section{Spherical coordinates in $D$ dimensions}

In obtaining explicit parallel one-forms fields for $\mathcal{M}_{in}$, it will be very convenient to introduce hyperspherical coordinates $(\theta_{1},..\theta_{j-1},\phi)$ in $S_{j}$, which are related to the cartesian coordinates $X_{j}$ in the internal space $R^{j+1}$, according to

\begin{equation} \label{coordinat}
X_{k} =\left\{\begin{array}{ccl}
r \cos \theta_{1} & \mbox{if} & k= 1\\
 && \\
r \cos \theta_{k}\, \Pi _{p=1}^{k-1} \sin \theta_{p} & \mbox{if} & k= 2, \ldots , j-1 \\
 && \\
r \sin \phi\, \Pi_{p=1}^{j-1} \sin \theta_{p}  & \mbox{if}  & k =j \\
 && \\
r \cos \phi\, \Pi_{p=1}^{j-1} \sin \theta_{p}  & \mbox{if}  & k =j+1 \\
\end{array}
\right.
\end{equation}
 Now, by inverting these equations, we obtain the spherical coordinates in terms of Cartesian coordinates, i.e., 
\begin{eqnarray}
\nonumber r^2&=&X_{1}^2+ \ldots +X_{j+1}^2\,,\\
\nonumber \theta_{k}&=&\arccos \left( \frac{X_{k}}{\sqrt{\sum_{i=k}^{j+1}X_{i}^2}} \right), \,\,\, k=1, \ldots , j-1\,,\\
\phi&=&\arctan \left( \frac{X_{j}}{X_{j+1}} \right)\,,
\end{eqnarray}
In order to change coordinates, we will need the Jacobian of the coordinate transformation in question. The derivatives of the above expressions with respect to $X_{i}$ are
\begin{eqnarray}
 \frac{\partial r}{\partial X_{k}}&=&\frac{X_{k}}{r}\,,\\
\nonumber \frac{\partial \theta_{l}}{\partial X_{k}}&=&\left\{\begin{array}{ccl}
-\frac{1}{\sqrt{\sum_{i=l}^{j+1}X_{i}^2-X_{l}^2}} \left( \delta_{kl}-\frac{X_{k}X_{l}}{\sum_{i=l}^{j+1}X_{i}^2} \right) & \mbox{if} & k\geq l\\
 && \\
0  & \mbox{if}  & k < l\,. \\
\end{array}
\right.
\end{eqnarray}

\section{On the remnant group of Lorentz transformations and the uniqueness of the vielbein field}\label{remnantgroup}

In this short section we discuss the remnant symmetries underlying $f(T)$ gravity, and their impact on the set of parallelizations admissible for a given spacetime. Details concerning the following exposition can be found in \cite{Nos2015}.

In general, under a Lorentz transformation of the vielbein $E^{a}\rightarrow E^{a'}=\Lambda^{a'}_{\ \ b}E^{b}$, the Weitzenb\"{o}ck invariant $T$ in $D$ spacetime dimensions transform as

\begin{equation}\label{Ltegfor}
T\rightarrow T'= T+e^{-1}d(\epsilon_{i_{1},...,i_{(D-2)},a,b}\,E^{i_{1}}...E^{i_{(D-2)}}\,\eta^{bc}\,\Lambda^{a}_{d}\,d\Lambda^{d}_{c})\,,
\end{equation}
where the wedge product $\wedge$ is understood. Note that $T$ is a scalar only under the \emph{global} Lorentz group ($d\Lambda^{d}_{c}=0$).

The remnant group $\mathcal{A}(E^a)$ of a given spacetime $(\mathcal{T}^{\star}\mathcal{M},E^a(x))$ is defined as the subgroup of $SO(1,D-1)$ under which $T$ becomes a Lorentz scalar, i.e., by demanding
\begin{equation}\label{defgropu}
d(\epsilon_{i_{1},...,i_{(D-2)},a,b}\,E^{i_{1}}...E^{i_{(D-2)}}\,\eta^{bc}\,\Lambda^{a}_{d}\,d\Lambda^{d}_{c})=0\,.
\end{equation}
If we consider infinitesimal Lorentz transformations
\begin{equation}\label{trnsfinfi}
\Lambda^{a}_{b}=\delta^{a}_{b}+\frac{1}{2}\sigma^{c\,d}(x)(M_{c\,d})^{a}_{b}+\mathcal{O}(\sigma^{2})\,,
\end{equation}
where $\sigma^{c\,d}(x)=-\sigma^{d\,c}(x)$ are the $D(D-1)/2$ parameters of the transformations, and
\begin{equation}\label{matricesm}
(M_{c\,d})^{a}_{b}=\delta^{a}_{c} \eta_{d\,b}-\delta^{a}_{d} \eta_{c\,b}\,,
\end{equation}
the term appearing in (\ref{defgropu}) results
\begin{equation}\label{dobleterm}
\Lambda^{a}_{d}\,d\Lambda^{d}_{c}\simeq -\frac{1}{2}d\sigma^{b\,d}(M_{b\,d})^{a}_{c}=\eta_{c\,b} d\sigma^{b\,a}\,.
\end{equation}
In this way, the condition (\ref{defgropu}) becomes
\begin{equation}\label{defgrupfirst}
\epsilon_{i_{1},...,i_{(D-2)},a,b}\,d(E^{i_{1}}\wedge...\wedge E^{i_{(D-2)}})\wedge d\sigma^{a\,b}=0\,.
\end{equation}
Recall that, in $D$ spacetime dimensions, we have $D-1$ boosts generators $K_{\alpha}=M_{0\alpha}$, and $\frac{1}{2}(D-1)(D-2)$ rotations $J_{\alpha}=-\frac{1}{2}\epsilon_{\alpha\beta\gamma}M^{\beta\gamma}$.

Not much can be said about the solutions of eq. (\ref{defgrupfirst}) in general. However, for the specific case under consideration, the structure of the parallel vector fields allow us to briefly explore some consequences of it. Regarding the vielbein components given in Eq. (\ref{esquema7D}), we immediately note that, due to the fact that $E^{0}=dt$, we have

\begin{equation}\label{0-caf}
d(E^{0}\wedge \tilde{\phi})=dt\wedge d\tilde{\phi}\,,
\end{equation}
for any of the $8-$forms $\tilde{\phi}$ which can be constructed by wedge products of $E^{i}$. Solutions of the Eq. (\ref{defgrupfirst}) will include, then, \emph{time dependent} corresponding Lorentz generators. For instance, if

\begin{equation}\label{0-cafbis}
\tilde{\phi}=E^{1}\wedge ... \wedge E^{8}\,,
\end{equation}
then a (time dependent) free rotation parameter $\sigma^{9\,10}(t)$ will solve Eq. (\ref{defgrupfirst}). This is so because $d\sigma^{9\,10}(t)\propto dt$, and then $d(E^{0}\wedge \tilde{\phi})\wedge d\sigma^{9\,10}=0$. Of course, many more time dependent rotations are allowed; actually, the full time dependent group of rotations about a certain axis is contained in (\ref{defgrupfirst}). It is possible to show that certain time-dependent Lorentz boost are also contained in $\mathcal{A}(E^a)$, see \cite{Nos2015}. This infinite set of allowed 1-form fields, each of them connected by remnant symmetries, are representative of the non uniqueness of the parallelization process of the cosmological manifold under consideration.

\section{$\mathcal{M}_{in}= S^3 \times S^2 \times S^2$}\label{anotherexample}

Here, as an example of ``non trivial'' internal space parallelization, we proceed to show a case in which the product topology of the internal dimensions is not constituted by parallelizable spheres. As a consequence of the non parallelizability of $S^2$, the one-form fields have not the block structure coming from the topological product, but instead, they will contain cross terms. Even though no global basis exist for $S^2$, it certainly exists for the three dimensional manifold $S^1\times S^2$, which is orientable. Explicit global fields for $\mathcal{T}^{\ast}(S^1\times S^2)$ are \cite{Brikel}

\begin{eqnarray} \label{camposens312}
  E^{1}(S^1\times S^2)&=&a_{0}\,X_{3}dx_{1}-a_{2}\,(X_{2}dX_{1}-X_{1}dX_{2})\,,\\ \notag
  E^{2}(S^1\times S^2)&=&a_{0}\,X_{2}dx_{1}+a_{2}\,(X_{3}dX_{1}-X_{1}dX_{3})\,,\\ \notag
  E^{3}(S^1\times S^2)&=&a_{0}\,X_{1}dx_{1}-a_{2}\,(X_{3}dX_{2}-X_{2}dX_{3})\,,\\ \notag
\end{eqnarray}
where the coordinates of $S^{1}\times R^3$ in which we are embedding $S^1\times S^2$, are $(x_{1},X_{1},X_{2},X_{3})$. Although we have not an $S^1$ in the internal space, the trick consists on using the periodic coordinate on it (here $x_{1}$), to ``rectify it'', and to think about it as one of the coordinates of the external space. Clearly, this process involves the inclusion of the spatial section of the $FRW^{4}$ space, so the block structure of the field is broken. In this way, a global basis for the entire eleven dimensional manifold $\mathcal{M}$ consists of two copies of the sort (\ref{camposens312}), one corresponding to $S^3$ (see Eq. (\ref{camposens3})), plus the temporal part and the remaining bulk dimension.

In spherical coordinates $(\theta_{1},\theta_{2},\phi_{1},\theta_{3},\phi_{2},\theta_{4},\phi_{3})$ we have the internal metric

\begin{equation}
  ds^2_{in} =a_{1}(t)^2\,d\Omega_3^{\,2}+a_{2}^2(t)\, d\Omega_{2,(1)}^{\,2}+a_{3}^2(t)\, d\Omega_{2,(2)}^{\,2}\,,
\end{equation}
where

\begin{eqnarray}
d\Omega_3^{\,2}&=&d\theta_{1}^2 + \sin\negmedspace^2\theta_{1}\, d\theta_{2}^2 +
  \sin\negmedspace^2\theta_{1} \sin\negmedspace^2\theta_{2}\, d\phi_{1}^2\,,\\
  d\Omega_{2,(1)}^{\,2}&=&d\theta_{3}^2 + \sin\negmedspace^2\theta_{3}\, d\phi_{2}^2 \,,\\
  d\Omega_{2,(2)}^{\,2}&=&d\theta_{4}^2 + \sin\negmedspace^2\theta_{4}\, d\phi_{3}^2\,.
\end{eqnarray}

The invariant $T$ is given by

\begin{equation}
   T= -2\Big(9H_{1}H_{0}+6H_{1}H_{2}+6H_{1}H_{3}+3H_{1}^{2}+6H_{0}H_{2}+6H_{0}H_{3}+3H_{0}^{2}+4H_{2}H_{3}+H_{2}^{2}+H_{3}^{2}-3a_{1}^{-2}-a_{2}^{-2}-a_{3}^{-2}\Big)\,.
\end{equation}
The $f(T)$ field equations are given by:
\begin{align}
  f + 2f'(2a_{3}^{-2}+2a_{2}^{-2}+6a_{1}^{-2}-T) = 16\pi G \rho \,,
\end{align}

\begin{multline}
  4 f'' \Bigl(\frac{3 H_1}{2}+H_0+H_2+H_3\Bigr) \dot{T} +
  4f' \Bigl(-\frac{3}{2} H_1 H_0+\frac{3}{2} \dot{H}_1+\frac{3}{2} H_1^2-H_0(H_2+H_3) \\ +\dot{H}_0+\dot{H}_2+H_2^2+\dot{H}_3+H_3^2+3a_{1}^{-2}+a_{2}^{-2}+a_{3}^{-2}-\frac{T}{2} \Bigr)+f = -16 \pi G p_{0} \,.
\end{multline}
\begin{multline}
  4 f'' \Bigl(H_1+\frac{3 H_0}{2}+H_2+H_3\Bigr) \dot{T} +
  4 f' \Bigl(-\frac{1}{2} H_1 (3 H_0+ 2 H_2+ 2 H_3)+\dot{H}_1 \\ +\frac{3}{2} \dot{H}_0+\frac{3}{2} H_0^2+\dot{H}_2+H_2^2+\dot{H}_3+H_3^2+2a_{1}^{-2}+a_{2}^{-2}+a_{3}^{-2}-\frac{T}{2} \Bigr)+f = -16 \pi G p_{1} \,.
\end{multline}

\begin{multline}
  2 f'' \Bigl(3 H_1+3 H_0+H_2+2 H_3\Bigr) \dot{T} +
  2 f' \Bigl(-3 H_1 H_2+3 \dot{H}_1+3 H_1^2+3 H_0(H_0-H_2) \\ +3 \dot{H}_0-2 H_2 H_3+\dot{H}_2+2 \dot{H}_3+2 H_3^2+6a_{1}^{-2}+a_{2}^{-2}+2a_{3}^{-2}-\frac{T}{2}\Bigr)+
  f = -16 \pi G p_{2} \,.
\end{multline}
\begin{multline}
  2 f'' \Bigl(3 H_1+3 H_0+2 H_2+H_3\Bigr) \dot{T} +
  2 f' \Bigl(-H_3 (3H_1+ 3H_0 +2 H_2)+3 \dot{H}_1+3 H_1^2 \\ +3 \dot{H}_0+3 H_0^2+2 \dot{H}_2+2 H_2^2+\dot{H}_3+6a_{1}^{-2}+2a_{2}^{-2}+a_{3}^{-2}-\frac{T}{2}\Bigr)+
   f = -16 \pi G p_{3} \,.
\end{multline}

\section{Explicit forms of functions $\mathcal{A}$ and $\mathcal{B}$}
\label{AppD}

\begin{multline}
\dot{H}_0=\biggl[b^2 f \left(21 \left(H_0+2 H_1\right) \left(H_0+8 H_1\right) f''-f'\right) \\ +6 f' \Bigl\{21 \Big(6 b^2 H_0^4+336 b^2 H_0 H_1^3+H_0 H_1 \left(59 b^2 H_0^2-46\right)
+3 H_1^2 \left(71 b^2 H_0^2-28\right)-8 H_0^2+196 b^2 H_1^4\Bigr) f''\\
+\left(14-3 b^2 H_0\left(3 H_0+7 H_1\right)\right) f'\Bigr\}\biggr]\Bigl/
\Bigl[18 b^2 f' \Big(f'-3 \Big(21 H_0 H_1
+4 H_0^2+14 H_1^2\Big) f''\Big)\Bigr]\,, 
\label{AppDEq1}
\end{multline}
\begin{multline}
    \dot{H}_1 = \biggl[6 f' \Big(-3 \Bigl(1029 b^2 H_0 H_1^3+21 H_1^2 (27 b^2 H_0^2-16) \\
    +4 H_0 H_1 (34 b^2 H_0^2-35)+4 H_0^2 \left(3 b^2 H_0^2-4\right)+686 b^2 H_1^4\Bigr) f''
     -\left(3 b^2 H_1 \left(3 H_0+7 H_1\right)+4\right) f'\Big)\\-b^2 f \left(3 \left(2 H_0+7 H_1\right) \left(H_0+8 H_1\right) f''(T)+f'(T)\right)\biggr]\Bigl/
     \Bigl[18 b^2 f'(T) \Big(f'-3 \Big(21 H_0 H_1
    +4 H_0^2+14 H_1^2\Big) f''\Big)\Bigr] \,,
\label{AppDEq2}
\end{multline}
with $f=T+\alpha T^2$, $f'=1 + 2\alpha T$, $f''=2\alpha$ and where $b$ is given by
\begin{equation}
b^2=-\frac{84 \alpha}{1+12 \alpha \left( H_0^2+7H_0 H_1 +7H_1^2 \right)+\sqrt{1+576 \alpha^2\left( H_0^2+7H_0 H_1 +7H_1^2 \right)^2 }} \,.
\end{equation}

\end{document}